# Molecular Dynamics Simulations

# of

# a Nucleosome and Free DNA

Thomas C. Bishop

Department of Environmental Health Sciences

Tulane University Health Sciences Center

1430 Tulane Avenue SL-29

New Orleans, Louisiana, 70112

ABBREVIATIONS:

    bps: basepairsteps

    bp: basepair

*To whom correspondence should be addressed

Phone: (504) 988-6203

Fax (504) 988-6428

Email: bishop@tulane.edu

**Abstract**

Nucleosomes organize the folding of DNA into chromatin and significantly influence transcription, replication, regulation and repair. All atom molecular dynamics simulations (10ns) of a nucleosome and of its 146 basepairs of DNA free in solution have been conducted. DNA helical parameters (Roll, Tilt, Twist, Shift, Slide, Rise) were extracted from each trajectory to compare the conformation, effective force constants, persistence length measures, and fluctuations of nucleosomal DNA to free DNA. A method for disassembling and reconstructing the conformation and dynamics of the nucleosome using Fourier analysis is presented. Results indicate that the superhelical path of DNA in the nucleosome is irregular. Long length variations in the conformation of nucleosomal DNA are identified other than those associated with helix repeat. These variations are required to create a proposed tetrasome conformation or to qualitatively reconstruct the 1.75 turns of the nucleosome's superhelix. Free DNA achieves enough bend and shear in solution to create an ideal nucleosome superhelix, but these deformations are not organized so the conformation is essentially linear. Reconstruction of free DNA using selected long wavelength variations in conformation can produce either a left-handed or a right-handed superhelix. DNA is less flexible in the nucleosome than when free in solution, however such measures are length scale dependent.

## Introduction

Eukaryotic genomes are organized in the cell nucleus as a hierarchy of folded structures by a protein-DNA complex called the nucleosome. Kornberg introduced the nucleosome hypothesis in 1974(Kornberg, 1974). Today, a number of high resolution x-ray crystallographic structures of the nucleosome are available from the protein databank. (For historical perspectives of the developments see the accounts in (Harp *et al*., 2000b; Khorasanizadeh, 2004; Kornberg & Lorch, 1999).) However, the organization of nucleosomes into chromatin and many of the structural and dynamical properties of the nucleosome itself are not well understood. Further characterization of the nucleosome is needed to understand its role in transcription, replication, recombination and repair.

The difficulties in characterizing the nucleosome are in part due to observations that the nucleosome can serve as an indiscriminate or specific modular entity. As an indiscriminant entity, the nucleosome directly organizes approximately 70% of the basepairs in the human genome, regardless of sequence identity(Noll, 1974). As a specific modular entity, some DNA sequences lead to nucleosome positioning(Widom, 2001). The key to crystallizing nucleosome core particles yielding high quality x-ray reflection data was strongly dependent upon the sequence of the DNA utilized for the sample preparation(Harp *et al*., 1996; Richmond *et al*., 1988). Thus quantifying differences between the structure and dynamics of a sequence of free DNA and its nucleosomal equivalent is a first step toward understanding DNA folding and chromatin structure. Such knowledge also impacts our understanding of how drugs, toxins, damaged DNA, or gene regulatory proteins alter the nucleosome and thus chromatin.

Here, molecular dynamics simulations are utilized to investigate the structure and dynamics of the 146bp sequence of DNA that is present in the majority of the nucleosome crystal structures (NCP146dna) (Luger *et al*., 1997). Simulations of this sequence as linear B-form DNA free in solution and in complex with histones in an engineered nucleosome complex (pdb entry 1KX3)(Davey *et al*., 2002) are reported.

These simulations are the first simulation of the nucleosome core particle based on a crystallographic structure and the longest length of linear DNA in explicit solvent simulated using all-atom molecular dynamics techniques. The simulation of DNA in solution contains approximately one persistence length of DNA and spans sufficient time and length scales to observe the initiation of DNA supercoiling at atomic resolution.

A pioneering effort to model the nucleosome core particle (Arents & Moudrianakis, 1993) and subject it to molecular dynamics simulations was reported previously, see www.t10.lanl.gov/angel/nucleo.html. In the present study, analysis focuses on how the conformation and dynamics of free DNA and nucleosomal DNA differ during 10ns molecular dynamics simulations. For this purpose, the DNA helical parameters (Roll, Tilt, Twist, Shift, Slide, Rise) (Dickerson & et al., 1989) as obtained by 3DNA(Lu & Olson, 2003) analysis of each trajectory are compared.

The comparison of helical parameters utilizes the same analysis techniques employed in (Lankas *et al*., 2003). The analysis provided dinucleotide force constants for every possible basepairstep assuming the energy is a harmonic function of all six helical parameters, i.e, the matrix K is determined for the expression $E = \tfrac{1}{2}(\mathbf{x} - \mathbf{x}_o)^T K(\mathbf{x} - \mathbf{x}_o)$ where $\mathbf{x}$ and $\mathbf{x}_o$ are six dimensional vectors representing the six helical parameters and the minimum energy conformation of each basepairstep, respectively. The theory is described briefly in (Lankas et al., 2003) and fully developed in (Gonzalez & Maddocks, 2001). Such analysis provides a bridge between the atomic scale of molecular dynamics and elastic rod models, e.g. (Coleman *et al.*, 2003; Manning *et al.*, 1996).

Here the technique is utilized as a means of comparing the structural and dynamical features of free DNA and nucleosomal DNA. Specifically the conformation, effective force constants, effective persistence length measures, and fluctuations of free DNA and nucleosomal DNA are compared. Fourier analysis of these results provides a method for investigating how the structure and dynamics varies on different length scales.

**Methods**

In this section the systems modeled, simulation techniques and analysis techniques are presented. The labels used to describe results are also presented.

*Systems Modeled:* Two systems were simulated. Each included a 9Å shell of explicit water, neutralizing ions and an excess of Na+ and Cl- ions corresponding to a concentration of 150mM NaCl.

*1kx3* is the label for the simulation of the nucleosome. It is based on the 2.0Å resolution protein databank entry 1kx3(Davey et al., 2002). The only heavy atoms missing are associated with the histone tails. No attempt was made to model the tails, as they are present in protein databank entry 1kx5(Davey et al., 2002). Entry 1kx3 was utilized because it allowed a smaller unit cell. The structure includes a histone octamer (H2A.1-H2B.2)(H3-H4)$_2$(H2A.1-H2B.2), 146bp, 13 Mn+ ions and 943 waters. The waters present in the x-ray structure were preserved when building the system, but the Mn+ was not. Instead 216 neutralizing Na+ ions were added with the addions command in tleap, followed by addition of a 9Å solvent layer. Subsequently 92 excess ion pairs were added using a VMD(Humphrey *et al.*, 1996) script that randomly chose two waters from bulk water and replaced one with $NA^+$ and one with $Cl^-$ for each ion pair. The choice of waters occurred iteratively. At each iteration bulk water was redefined as any water greater than 5Å from protein, nucleic or ion. (This VMD script is freely available upon request.) The final system contained 134,335 atoms (112,116 water, 12,535 protein, 9,284 DNA, 308$Na^+$, 92$Cl^-$) with zero net charge inside a cell of approximately 123Å x 123Å x 88Å as measured at the end of equilibration. These dimensions varied by less than 1% during the simulation.

The initial system was subjected to 500 steps of minimization during which any atom was allowed to move a maximum of 0.1Å per step. Minimization beyond this point was not a necessary prerequisite for equilibration. The system was equilibrated in two stages. Both stages utilized NAMD's Berendsen pressure coupling scheme (relaxation time 500fs, target pressure 1.01325atm, compressibility 4.57E-5), PME electrostatic calculations at every step (grid size 128Å x 128Å x 128Å, 4th order interpolation, Ewald coefficient 0.31), NAMD's C1 switching from 8Å to 10Å, pairlist cut-off of 12Å, 1-4scaling of 0.8333, the SETTLE algorithm applied to all hydrogen, and a time step of 2fs. In stage one, 150ps were computed with a temperature reassignment every picosecond that began at 0K, was incremented by 50K for each reassignment, and achieved a maximum reassignment temperature of 300K. During the second stage 150ps were conducted with NAMD's Langevin temperature control turned on (target temperature 300K, damping coefficient $0.2ps^{-1}$) and the temperature reassignment turned off. After equilibration 10.0ns of NTP dynamics were computed using the simulation protocol described in the Simulation Techniques section. Only the NTP dynamics are considered for analysis.

*dna* is the label for the simulation of 146bp of DNA free in solution. The DNA had the same sequence as in *1kx3*. The model was constructed to have a linear B-form conformation using 3DNA(Lu & Olson, 2003) and to fit in a periodic cell of length 485Å with exactly 14 helical turns, (146bps/14 =10.43bps/turn = 34.52°/bps). This is overtwisted compared to ideal B-form DNA, ~10.5bps/turn, but has the same average twist value reported for nucleosomal DNA(Richmond & Davey, 2003). The integral number of turns enabled the two ends of the DNA to be connected to create a periodic chemical structure (PCS). To create the PCS the phosphate of the first residue of the A chain was bonded to the O3' of the last residue of the A chain. Likewise the last phosphate of the B chain was bonded to the first O3' of the B chain. The use of a PCS negates end effects, as there are no ends, and introduces a topological restraint as occurs for circular DNA:

$$Lk = Tw + Wr \qquad\qquad 1$$

where Lk is Linking number, Tw is the total Twist and Wr is Writhe. Lk is the sum of the signed crossings of one DNA strand over the other, and Writhe is the sum of the signed crossings of the DNA over itself. Writhe requires Bend. Since the initial model was constructed without Bend, $Wr_o = 0$ and $Lk_o = Tw_o = 14$ right-handed turns. Unless the covalent bonds are broken $\Delta Lk = 0$ in *dna* so any change in Tw during the simulation introduces Wr, i.e. $\Delta Tw = -\Delta Wr$. If Bend develops in *dna*, the PCS introduce a tension in the DNA since the end-to-end length of the DNA becomes shorter than the periodic cell. For a persistence length of DNA (~150bp) one expects the ends of the DNA to differ in orientation by 1 radian (57°) due to thermal motion, (Marko, 2004), yet the PCS as implemented here requires the difference to be zero. Experimentally, a force of ~10pN must be applied to straighten out thermal fluctuations in DNA before the DNA is actually stretched(Leger *et al*., 1999). Thus, as implemented here, the PCS introduces tension in DNA that in turn causes the pressure regulation algorithm to shorten the unit cell along the axis of the DNA as a means of negating the tension.

*dna*, like *1kx3*, had a solvent shell extending 9Å beyond the solute. The shell included neutralizing ions, and an excess of $Na^+$ and $Cl^-$. The final system for *dna* consisted of 93,319 atoms (83,607 waters, 67Cl-, and 359Na+) and measured 44Å x 44Å x 480Å at the end of equilibration. The long axis gradually shortened by 3% over the course of the simulation while each short axis lengthened by 1.6% such that the volume varied by less than 0.3%. The short axes of the unit cell were constrained to vary isotropically in NAMD.

*Simulation Techniques:* NAMD2 version 2.5 (Kale *et al*., 1999) was used for the compute engine because its implementation of periodic boundary conditions is compatible with the PCS described above. Simulations were conducted on either an 8 processor SGI O300-R14000 or 2 to 3 nodes of a Rocks Linux cluster with 2-AMD Opteron-1.9Ghz processors/node. All structures were built using leaprc.ff99 in Amber7 which loads the parm99.dat force field parameter set and utilizes the Cornel 1994 topologies. It was stated in (Lankas et al., 2003) that "...Parm99 does not seem to show proper A-B equilibrium or bending dynamics (data not shown)". However the results presented in Table I are within the range of values obtained in (Lankas *et al*., 2002; Lankas *et al*., 2000). 10ns of NTP dynamics were computed using the same simulation options as for equilibration except but with a pressure regulation time constant was 5ps. An additional 100ps of dynamics in the NVE ensemble (temperature and pressure regulation turned off) were computed for each system using the restart files at t=1.0ns from the NTP dynamics. Analysis of these trajectories indicates a relative fluctuation in the total energy of $10^{-6}$, but no drift. The fluctuation is a rigorous measure of the order of accuracy of the numerical integration as implemented in NAMD with the above compute options.

*Analysis Techniques:* Snapshots from the 10ns of NTP dynamics were saved every 500fs yielding 20,000 snapshots for each simulation. 3DNA was utilized to determine DNA helical parameters (Roll, Tilt, Twist, Shift, Slide, Rise) for each basepairstep for all

20,000 snapshots for each simulation. Three additional quantities, Shear, Bend and Distance are calculated:

$$Dist = \sqrt{(Shear)^2 + (Rise)^2}$$
$$Shear = \sqrt{(Shift)^2 + (Slide)^2} \quad\quad 2$$
$$Bend = \sqrt{(Roll)^2 + (Tilt)^2}$$

The analysis programs and scripts utilized in (Lankas et al., 2003) were made available for the analysis presented here. As provided the routines calculated Roll, Tilt, Twist, Shift, Slide, Rise and Dist values from unit directors determined by 3DNA for each basepairstep. Force constant and persistence length data in Table I were determined using the covariance matrix inversion technique described in (Lankas et al., 2003). FFTW(Frigo & Johnson, 1998) was incorporated into the programs to calculate discrete Fourier transforms of the helical parameters as a function of basepairstep. There are 145 basepairsteps so the Fourier transform index, i, ranges from 1-72 corresponding to a wavelength of 145/i in terms of basepairsteps or 145*<Dist>/i in Ångstrom. This approach enables three different length scales to be defined: 1) **Long** i=[1,14], corresponding to wavelengths greater than one helix repeat 2) **Intermediate** i = [15-48], between one helix repeat and 3 basepairsteps and 3) **Short** i = [49-72], less than 3 basepairsteps. The Fourier transform data as plotted in Figure 2 is not normalized, thus the vertical axes represent the amplitude of the oscillation relative to the average values reported in Table I. The Fourier spectra are normalized to determine the percent contribution from each length scale as reported in Table II.

**Results and Discussion**

In this section overall conformational and dynamical properties of DNA are considered followed by a Fourier analysis. The overall properties are each calculated as an arithmetic mean of the values obtained for all basepairsteps. Basic measures of conformation, namely percent of Watson-Crick pairing and B-form like conformation, are presented first to assess the integrity of the DNA in each simulation.

**Pairing and Stacking**

For *dna* greater than 99% of all snapshots possess complete, canonical Watson-Crick basepairing as determined by 3DNA. The PCS allows even the bases at each "end" of the DNA to remain properly paired and stacked. The PCS utilized in *dna* is proposed to offer a more accurate representation of DNA within its genomic context than free end simulations since PCS removes ends effects and introduces constraints that limit conformational freedom of the DNA. Fraying and improper stacking of the DNA clearly affects the terminal bases in *1kx3*, which did not have a PCS.

During the first nanosecond of *1kx3* as many as 5 base steps are identified as unpaired or mismatched and less than 10% of the snapshots exhibit complete pairing. Over the course of the first 4.0ns the basepairing improves so that after 4.0ns greater than 90% of all of the remaining snapshots exhibit complete, canonical Watson-Crick pairing and a maximum of 2 unpaired basepairsteps for any given snapshot.

To further characterize the conformation of the DNA, the number of basepairs identified as "B-form" using the default definition in 3DNA is determined. For *dna* a maximum of 136 and a minimum of 92 of the 146 basepairs are identified as canonical B-Form for any single time step. For any given snapshot of *dna* 118 basepair steps are likely to be identified as having a B-form like conformation. The *dna* basepairs at positions 72-74, corresponding to the dyad in *1kx3*, exhibit the highest frequency of B-form like conformations, and positions 119-120 exhibit a higher than average frequency of B-form conformation. Basepairs 1 and 146 exhibit an average frequency of B-form like conformations, again indicating that the PCS successfully eliminates end effects.

In *1kx3* a maximum of 127 basepairs are classified as B-form for any single snapshot in which all bases are fully paired. The minimum is 85 and the average is 108. Basepairs 119-20, which are considerably stressed in *1kx3*, exhibit the lowest frequency of B-form conformation while basepairs near the dyad maintain a higher than average frequency of B-form conformation.

**Properties of DNA**

Table I summarizes the properties of DNA determined as the arithmetic mean of the time averaged values obtained for all basepairsteps. The conformation of DNA is more distorted in *1kx3* than in *dna* as determined by any measure, e.g. Bend, Shear, Dist, or Twist. The flexibility of DNA as measured by standard deviations, force constants, or persistence lengths indicates a mild stiffening of the DNA. A comparison of the average conformational data is considered first.

**Conformation**

The time and length scales for *dna* are sufficient to observe significant deviations from a linear conformation. In fact enough Bend (7.0°/basestep) and Shear (0.94Å/basestep) develop in *dna* to create an ideal nucleosome superhelix. Such a superhelix can be created with a Bend of 360°/turn*1.75turns/145bps ~ 4.3°/bps to obtain the required curvature and a Shear of 28Å/turn*1.75turns/145bps ~ 0.34Å/bps to obtain the required pitch. The data from *1kx3* in Table I indicate that an actual nucleosome has an even greater excess of Bend (8.8°/bps) and Shear(1.0Å/bps). The path of DNA around the nucleosome is not a minimal superhelical path. Such irregularities provide a sufficiently loose wrapping of the DNA around the histones to accommodate thermal motions and conformational variations arising from the sequence specific properties of DNA or DNA binding proteins. A minimal superhelical path would not accommodate any variation without fundamentally altering the histone-dna interactions or severely distorting the local structure of the DNA.

Of course, Bend and Shear must be properly distributed along the DNA to obtain the nucleosome's superhelix, so consideration of only the magnitude of Bend and Shear can be misleading. For this purpose Writhe as a measure of DNA topology can be determined in case of *dna* and compared to an assumed Writhe of –1.75 turns for *1kx3*. The initial conformation used in *dna* had excess Twist (34.52°/bp or 14 turns) compared to the average of 34.3°/bp during dynamics. The latter is equivalent to the accepted helix repeat for B-form DNA of 10.5bp/turn. This relaxation created a positive Writhe (average 1/12 turns) because of the topological constraints imposed by the PCS. Thus even though the average total Bend in *dna* is large (1017° ~ 2.8 turns), it creates only 1/12 of a turn of Writhe. Writhe in *dna* is of the opposite sign and considerably lower in magnitude than the Writhe in the nucleosome (-1.75 turns). The handedness of the Writhe in *dna* is a consequence of relaxation from an initially overtwisted conformation. For *dna* to relax from a linear conformation to a conformation with a Wr of –1.75 turns and an average Twist equivalent to that observed in *1kx3* (34.76°/bp), the initial conformation in *dna* would have needed to be significantly underwound (Twist = 30.42°/bp). Experimental measures indicate that the change in Twist upon nucleosome formation does not account for the changes in Lk and Wr, the so-called Linking number paradox, reviewed in (Prunell, 1998). Twist is not constant in either free or nucleosomal DNA, see Figure 2, so caution should be exercised in using average Twist values, rather than an explicit sum, to determine the cumulative value of Twist required in Equation 1. Similar caution should be exercised in using a presumed helical geometry for the nucleosome to convert between local and intrinsic measures of DNA helicity because the nucleosome's superhelix is not smooth.

The development of Bend in *dna* reduces the end to end distance of the DNA creating tension. In this regard coupling of the PCS with the pressure regulation algorithm in NAMD favors the development of Writhe, Twist relaxation, and shortening of the unit cell length. Deliberate manipulation of the cell length, introducing a tension, or modifying the pressure calculation is required to counter this interaction between the PCS and pressure regulation. Similar behavior is expected for a PCS simulation of DNA with an initially underwound conformation but due to the topological constraints it will develop negative Writhe as the Twist relaxes to a higher value.

**Flexibility**

The average force constants for DNA obtained from either *dna* or *1kx3* are within the range of values reported in (Lankas et al., 2003). The average over all basepairsteps of the ratio of force constants, i.e. <Tilt/Roll>, is 1.56 in *dna* and 1.49 in *1kx3*; for <Slide/Shift> the values are 1.74 and 1.55 respectively. Thus, anisotropy associated with Bend is reduced by ~5% upon nucleosome formation and for Shear the reduction is more than 10%. These values for *dna* agree favorably with the average values of <Tilt/Roll> = 1.64 and <Slide/Shift> = 1.74 obtained by consideration of all possible combinations in (Lankas et al., 2003)

A comparison of the force constants obtained from *1kx3* to those from *dna* indicate that the histones tend to stiffen the DNA, most notably in terms of Shift and Slide. The exceptions are Rise and Tilt, which are softer in *1kx3*. If values of standard deviation are utilized as a measure of flexibility the data reported in Table I for Roll indicate what may initially be considered anomalous behavior. For a single variable, the persistence length ($p$), force constant ($f$) and fluctuation ($sd$) are related as follows (conversion factors omitted):

$$p = Dist * f = Dist / sd^2 \qquad 3$$

Thus the arithmetic mean of the force constants should be compared to the harmonic mean of the square fluctuation values, not the arithmetic mean values listed in Table I. The above relations are also a reminder of the role of the interbasepair distance, *Dist*, in determining persistence length. Since <*Dist*> in *1kx3* is 3% greater than that obtained for *dna* the effective persistence lengths obtained from *1kx3* will be 3% higher than those obtained from *dna* if the fluctuations are identical. Such is the case for Roll and Tilt; the difference in fluctuation values between *1kx3* and *dna* are not significant, nonetheless the persistence length measures for Roll, Tilt and therefore Bend in *1kx3* are each ~3% longer. For Twist, the higher value of *Dist* in *1kx3* accounts for most but not all, of the increase in persistence length. Persistence length properly measures the combined effects of segment length and inter-segment rotational motion, which are not captured by consideration of either force constants or fluctuations alone.

**Thermal Accessibility**

Fluctuations in the conformation of DNA as observed in *dna*, are compared to the average conformation observed in *1kx3* to determine whether or not the conformation of each basepairstep in the nucleosome can be achieved by thermal fluctuations. For this purpose the "envelope of allowed conformations" is defined at each basepairstep as the average value of the helical parameter +/- its standard deviation as obtained from *dna*. Figure 1 indicates that the conformation of DNA in *1kx3* is mostly within the envelope of allowed conformations. Thus deforming linear DNA into a nucleosomal superhelix only requires the histones to actively deform DNA at the isolated locations where the conformation in *1kx3* is outside the envelope. In case of Roll this occurs once every helix repeat. For the other helical parameters the spacing does not exhibit such a simple pattern.

Close inspection of Figure 1 indicates that the distribution of Roll in *1kx3* is most negative near the dyad and gradually increases to positive values away from the dyad. A parabolic fit to the data indicates a variation of approximately 3°. Similarly Slide and Rise vary by –0.6Å and –0.1Å respectively but the curvature is opposite (as indicated by the minus sign). The parabolic variation in Roll increases Bend near the dyad by approximately 20% and if properly phased would create an egg or pear shaped profile for the nucleosome. Such a conformation is not observed in the *1kx3*, but is observed in electron cryomicroscopy of chromatin fragments (Bednar *et al*., 1998). Theoretically, condensed chromatin corresponds to second order superhelical folding as described by elliptic functions and integrals; isolated nucleosomes correspond to first order superhelical folding as described by trigonometric functions (Shi and Hearst 1994; Shi, Borovik et al. 1995). The parabolic variation in helical parameters and tripartite organization of the histone octamer may reflect a built-in capability of the nucleosome for realizing higher order superhelical folding with non-circular cross-sections(Bishop & Hearst, 1998), regardless of the specific organization of nucleosomes within condensed chromatin.

**Fourier Analysis**

The periodic distribution of Roll as a function of basepairstep in *1kx3* is readily explained by the fact that the force constant for Roll is less than the one for Tilt. In the nucleosome,

DNA achieves most Bend by means of Roll. An elastic rod with anisotropic bend stiffness will also bend most where it is easiest to bend. Thus the distribution of Roll in the nucleosome varies sinusoidally with the cumulative value of Twist, achieving a maximal and minimal value for each successive turn. It is phased to achieve a maximum at the dyad. Below, Fourier analysis is used to further investigate the distribution of helical parameters as a function of length, measured in basepairsteps. A similar analysis is applied to the standard deviations obtained for each helical parameter.

Fourier analysis of the helical parameters indicates that in all cases the percent contribution of long wavelength modes is from 1 to 9% greater in *1kx3* than in *dna*, while the percent contribution of intermediate modes is from 2 to 13% less in *1kx3* than in *dna*, see Table II. In some cases the short wavelength contribution is higher for *1kx3* than *dna*, for others it is lower. Thus the conformation of DNA in the nucleosome is smoother (tends to vary on longer length scales) than free DNA even though it is more deformed.

Fourier analysis of the standard deviations indicates that 41-52% of the variation in the data from *1kx3* and *dna* comes from intermediate wavelength contributions, 20-40% by short and 14-35% by long, see Table II. Variations in DNA flexibility (as measured by standard deviation) are thus accounted for by intermediate, then short then long wavelength variations. The difference values for the standard deviation data listed in Table 2 indicate flexibility tends to vary on longer length scales in *1kx3* than in *dna*. The long length scale differences are all positive. Short and intermediate differences are mostly negative. Thus nucleosome formation changes the dynamic character of DNA by redistributing the flexibility over different length scales, even though the overall flexibility does not change significantly.

The contributions from each Fourier mode are presented in Figure 2 as the difference between Fourier data obtained from *1kx3* and that obtained from *dna*. A positive value indicates that *1kx3* possesses a larger amplitude oscillation for the indicated wavelength than the corresponding mode in *dna*. The opposite occurs for negative values. Maxima at index 14 indicate the importance of this mode to the structure and dynamics of the *1kx3*. This mode corresponds to local minima in *dna*. Local maxima occur at index 11 and 12 in *dna*.

Below, the helical parameter data is utilized to reconstruct models of the DNA in each simulation using the average conformation in Table 1 plus selected modes from the Fourier analysis. All six helical parameters are used for the model building. The standard deviation data is not. Analysis of such fluctuations will be considered elsewhere. The models in Figure 3 represent the effects of increasing the long wavelength variation for each reconstruction. This progression therefore reflects the structural organization that occurs during nucleosome formation.

**Short Wavelengths**

As may be expected short wavelength ($< 3.0$bps) variations in the conformation of the DNA do not yield significant deviations from a linear conformation. Rapid oscillations in conformation cancel each other before accumulating so that the average value of each helical parameter determines the overall conformation. The result is that the model reconstructed using only short wavelength data agrees with Table I: the model for *1kx3* is longer, has more Twist and less Roll than the model for *dna*. In either case the model has a nearly linear B-form conformation, in agreement with the analysis of stacking and pairing. Variations introduced and suppressed by the histones on this length scale, see Figure 2, are the most sensitive to sequence specific variations in DNA and therefore likely play a role in nucleosome positioning.

**Intermediate Wavelengths**

In case of *dna*, utilizing only the intermediate wavelength (9.7 to 3.0bps) variations yields a left-handed superhelix. This is surprising because the superhelix is right handed when all modes are utilized, see Figure 3. The longest wavelength (9.7bps), mode 15, is sufficient to qualitatively reproduce this left-handed superhelix.

In case of *1kx3*, using only intermediate wavelength variations creates approximately one turn of an extended left-handed superhelix. Again, mode 15, is sufficient to qualitatively reproduce this extended superhelical conformation.

**Long Wavelengths**

Since the DNA sequence used in this study has a strong nucleosome positioning signal, it is natural to expect maxima at index 14 rather than minima in the data from *dna*. Maxima in *dna* occur at indices 11 and 12, producing a variation in the conformation of

DNA on a length scale that is longer than the helix repeat, approximately 40-44Å in terms of contour length. Since Rise in *1kx3* is approximately 3% more than in *dna*, index 14 in *1kx3* corresponds to approximately the same physical distance as index 11 in *dna*. Reconstruction of DNA using only mode 11 or 12 or modes 11 and 12 creates a right handed superhelix. Mode 12 makes the greater contribution to this conformation. Using all long wavelength modes, 1-14, introduces more writhe but does not qualitatively change the model.

Since only one sequence of DNA has been used in this study, the present simulations are not sufficient to determine whether the long length variations observed in *dna* are intrinsic long length properties of DNA or a property of this particular sequence.

Each helical parameter for *1kx3* has a maxima at index 14, see Figure 2, corresponding to the average Twist of DNA reported in Table 1 ($360°*14/145bps = 34.8°/bps$) and the 14 points of contact between DNA and the histones. Reconstructing DNA using only mode 14 creates 1.5 turns of the superhelix but without any pitch; the DNA self-intersects. For each helical parameter mode 14 contributes less than 5% to the total power spectra, but this single component of the spectra creates most of the Bend necessary to wrap the DNA around the histone core. This result, which uses all six helical parameters, differs from an analysis that considers only Roll and Tilt variations corresponding to mode 14 (Richmond & Davey, 2003). It therefore, underscores the importance of accounting for all six helical parameters rather than just ones producing Bend.

Combining modes 14 and 15 for *1kx3* yields a superhelix with the proper pitch and radius for the central ¾ superhelical turn near the dyad. The central loop is proposed as an approximate structure for the tetrasome's superhelix. Using modes 14-72, adds a small bulge that alters the entry-exit of the DNA tails into the central loop but leaves the central loop largely intact, see Figure 3c. Adding mode 13 significantly changes the model. It possesses a full loop, but the loop is located at one end rather than centrally. Sequentially adding modes 12, 11, down to 2 gradually changes the model. A proper nucleosome superhelix requires inclusion of the longest wavelength variations, i.e. mode 1, indicating the nucleosome has a significant anti-symmetric organization with respect to the dyad. A realistic superhelix can be achieved using only long and intermediate wavelength modes, i.e, modes 1-48 are a close approximation to Figure 3d.

In case of *1kx3* an analysis of phasing indicates that for Roll modes 7,14,28, and 56 are largely in phase with each other but anti-phased with the standard deviation, so that Roll is maximal at the dyad while standard deviation is minimal. This agrees with the phase of Roll reported in (Richmond & Davey, 2003) and the observation that the lowest temperature factors occur at the minor groove-in positions in the x-ray data(Harp *et al.*, 2000a). The standard deviation data for Tilt and for Slide have minimal contribution from mode 14, even though the conformation has a significant contribution from mode 14. The primary long wavelength contributions to standard deviation of Tilt and Slide are modes 12 and 13, respectively. In this case structure and dynamics can be separated, as is done in the analysis of the bend-shear dynamics of an elastic rod relative to some intrinsic conformation(Bishop *et al.*, 2004).

**Conclusion**

The conformational and dynamical properties of a sequence of DNA free in solution have been compared to the properties of the same sequence in complex with a histone octamer, as observed during 10ns molecular dynamics simulations. Analysis of the interbasepair helical parameters, (Roll, Tilt, Twist, Shift, Slide, Rise), indicates: 1) that the superhelical conformation of DNA in the nucleosome is irregular at the base-pair level and largely within the range conformations thermally accessible to free DNA, 2) that free DNA exhibits sufficient Bend and Shear in solution to create an ideal nucleosome superhelix, 3) that DNA is less flexible in the nucleosome that when free in solution, but this result depends on the length scale of the measurement and 4) that the histone core shifts the distribution of fluctuations in free DNA toward longer wavelengths upon nucleosome formation. A long length scale variation of the structure of free DNA of 40-44Å is also identified that may or may not be specific to the sequence used in this study.

Fourier analysis of the helical parameters as a function of length, measured in basepairsteps, allows the nucleosome or free DNA to be reconstructed from selected modes. There are two wavelengths governing the large-scale conformation of a proposed tetrasome: 145/14 ~ 10.4 and 145/15 ~ 9.7 basepairsteps. The least common multiple is 145 basepairsteps suggesting a rational basis for the length of DNA in a complete nucleosome. There are two long wavelength modes dominating the conformation of DNA free in solution, 145/11 ~ 13.2 and 145/12 ~ 12.1 basepairsteps. Again, the least common multiple is 145 basepairsteps. If the results are unique to this sequence, it suggests a unique nucleosome positioning signal exists. If the results are valid for any sequence they point to 145basepairsteps as a natural length of DNA for the nucleosome core particle. Interestingly, Fourier analysis has enabled identification of long wavelength variations in free DNA that produce left handed and right handed superhelix deformations.

Fourier analysis also indicates that the histones affect a redistribution of dynamical properties from shorter to longer length scales. Thus, as in the case of determining the handedness of the superhelix, the length scale over which flexibility is measured becomes important. This may explain why an increase in DNA overall flexibility upon nucleosome formation is reported elsewhere(Sivolob & Prunell, 2004). Some modes lead to large conformational changes in the structure of the nucleosome while others have minimal effect.

The conformation of nucleosomal DNA is found to be within the range of conformations accessible by thermal fluctuations of free DNA, at least for this sequence. Roll is the notable exception. The highly negative values of Roll that occur every helix repeat are ~5° outside of the range of the thermally accessible values. The positive values of Roll observed in the nucleosome can largely be achieved by thermal fluctuations because of the parabolic variation of Roll as a function of basepairstep. Given these observations it is reasonable to expect that once a histone octamer or tetramer contacts DNA, two complete turns of the DNA (one helix repeat in each direction) will wrap around the histones primarily due to thermal motion. In order to continue wrapping around the histones requires energy, which is readily supplied by the favorable contacts between the histones and the DNA phosphates.  As suggested by a colleague, this is a "fly-paper" binding motif, in which attractive interactions between two binding sites, one of which is tethered to a site in the neighborhood of the other, can be realized by thermal motions. Wrapping of the polymer around the substrate occurs if there is a small bias in favor of binding. Once several contacts are established significant energy is required to alter the interface because all of the contacts are coupled. Likewise, if the bias for binding is altered to an unfavorable interaction, unwrapping occurs and more readily begins at the ends.   Applied forces or torques (e.g. atomic tweezers or the cell machinery), electrolytic changes, or chemical modification of the histones or DNA can affect such a change in bias.

**Acknowledgments**

Equipment was provided by grants to T.C.B. through NSF Cooperative Agreement with the Louisiana Board of Regents (EPS-9720652 NSF/LEQSF 1996-1998-SI-JFAP-04) and Tulane University Wall Funds.  I would also like to thank members of the Center for Computational Sciences for their support including use of the Center's grid computing resources, and Philip Lankas for making his helical analysis scripts available.  Dave Maag, of the Center for Bioenvironmental Research, offered the fly-paper analogy.

|  | *dna* | | | | *1kx3* | | | |
|---|---|---|---|---|---|---|---|---|
| Helical Parameter | <Avg> | <Std> | force constant (kcal/mol) | Persistence Length (nm) | <Avg> | <Std> | force constant (kcal/mol) | Persistence Length (nm) |
| Roll(°) | 2.43 | 5.23 | 0.0261 | 48.0 | 0.94 | 5.27 | 0.0273 | 51.5 |
| Tilt(°) | 0.00 | 4.30 | 0.0399 | 73.7 | -0.06 | 4.44 | 0.0393 | 74.5 |
| Twist(°) | 34.3 | 5.21 | 0.0375 | 69.4 | 34.76 | 4.57 | 0.0492 | 93.7 |
| Bend(°) | 7.01 | 3.66 | | 57.5 | 8.83 | 4.15 | | 60.0 |
| Shift(Å) | 0.00 | 0.66 | 1.74 | | 0.03 | 0.59 | 2.38 | |
| Slide(Å) | -0.25 | 0.57 | 2.98 | | -0.09 | 0.52 | 3.55 | |
| Shear(Å) | 0.94 | 0.43 | | | 1.02 | 0.43 | | |
| Rise(Å) | 3.32 | 0.30 | 9.20 | | 3.36 | 0.31 | 9.10 | |
| Dist(Å) | 3.36 | | | | 3.46 | | | |

Table I: DNA conformation and dynamics.

Arithmetic mean of the 145 values obtained from *1kx3* and *dna* for observables associated with each helical parameter. See text for further details.

|        |      | Rise    |     | Roll    |     | Shift   |     | Slide   |     | Tilt    |     | Twist   |     |
|--------|------|---------|-----|---------|-----|---------|-----|---------|-----|---------|-----|---------|-----|
| Short  | *dna*  | 32%     | 33% | 36%     | 40% | 32%     | 22% | 32%     | 33% | 34%     | 35% | 46%     | 24% |
|        | *1kx3* | 36%     | 28% | 35%     | 32% | 43%     | 20% | 27%     | 35% | 36%     | 29% | 46%     | 22% |
|        |      | 4%      | -5% | -1%     | -8% | 11%     | -2% | -4%     | 1%  | 2%      | -5% | -1%     | -1% |
| Inter. | *dna*  | 59%     | 48% | 48%     | 41% | 56%     | 49% | 48%     | 48% | 50%     | 51% | 45%     | 52% |
|        | *1kx3* | 48%     | 45% | 43%     | 43% | 43%     | 46% | 44%     | 45% | 47%     | 51% | 43%     | 51% |
|        |      | -11%    | -3% | -6%     | 2%  | -13%    | -4% | -4%     | -3% | -3%     | 0%  | -2%     | -2% |
| Long   | *dna*  | 10%     | 19% | 15%     | 19% | 12%     | 28% | 20%     | 19% | 16%     | 14% | 9%      | 24% |
|        | *1kx3* | 17%     | 27% | 22%     | 25% | 13%     | 35% | 29%     | 20% | 18%     | 20% | 11%     | 27% |
|        |      | 7%      | 8%  | 7%      | 6%  | 2%      | 6%  | 9%      | 2%  | 1%      | 6%  | 2%      | 3%  |

Table II: Percent Contribution from Different Length Scales.

The percent contributions from three length scales to the normalized Fourier spectra are listed. The length scales are: Long (modes [1-14] ~10.4 – 145bps), Intermediate (modes [15-48] ~ 9.7-3.0bps), and Short (modes [49-72]~ 2.0bps-2.9bps). For each length scale the values obtained from *dna* are listed on top, from *1kx3* in the middle, and the difference at the bottom. The first value corresponds to the spectra of the helical parameter; the second corresponds to the spectra of the standard deviation of the helical parameter.

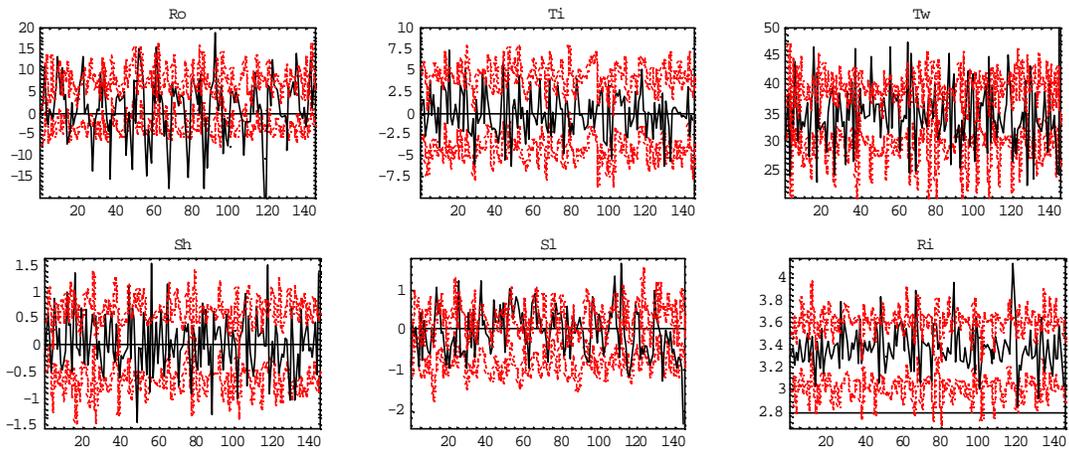

Figure 1. Thermal Accessibility.

Solid black lines represent the average helical parameter values determined from *1kx3*. The dotted red lines represent the envelope of allowed conformations observed during *dna*. At each point the envelope is the average value +/- one standard deviation. With the exception of Roll the helical parameters for *1kx3* are largely within the envelope of conformations accessible to DNA by thermal motion. Data corresponds to average and standard deviation as determined from simulation time 4.5ns-10.0ns. Roll, Tilt, and Twist values are in degrees. Shift, Slide, and Rise are in Å.

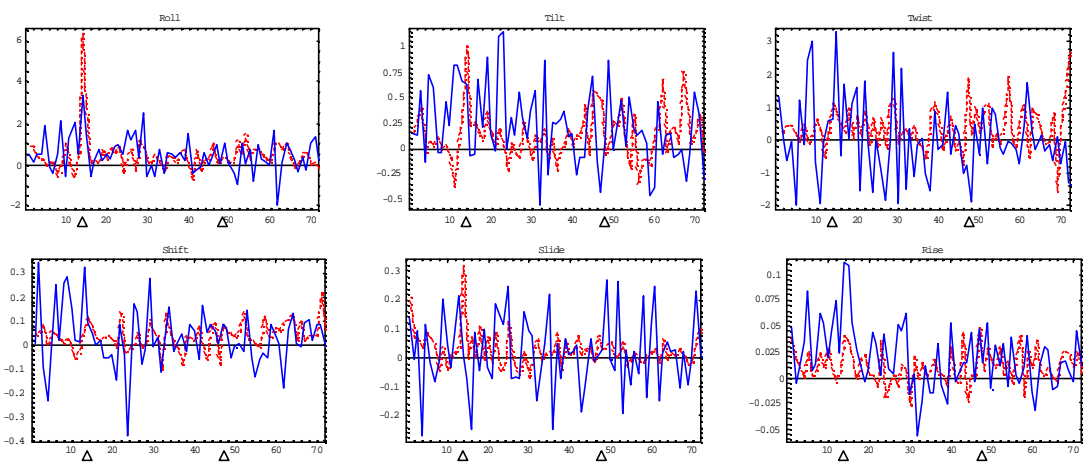

Figure 2: Fourier analysis of Helical Parameters and Standard Deviations.

The solid blue lines are differences *1kx3 - dna* between the Fourier transforms of the helical parameters. The dotted red lines are the differences *1kx3 - dna* of the Fourier transforms of the standard deviations. The data is not normalized so differences correspond to actual differences in amplitude. Negative values indicate the amplitude of the mode is greater in *dna* than *1kx3*. The vertical axes for Roll, Tilt and Twist are in degrees. Shift, Slide, Rise are in Å. The standard deviation data has been multiplied by a factor of 10.0. The horizontal axis in each case represents the Fourier index, 1< i <72, corresponding to wavelength 145/i in basepairsteps. Triangles mark boundaries between Long, Intermediate and Short wavelength index values

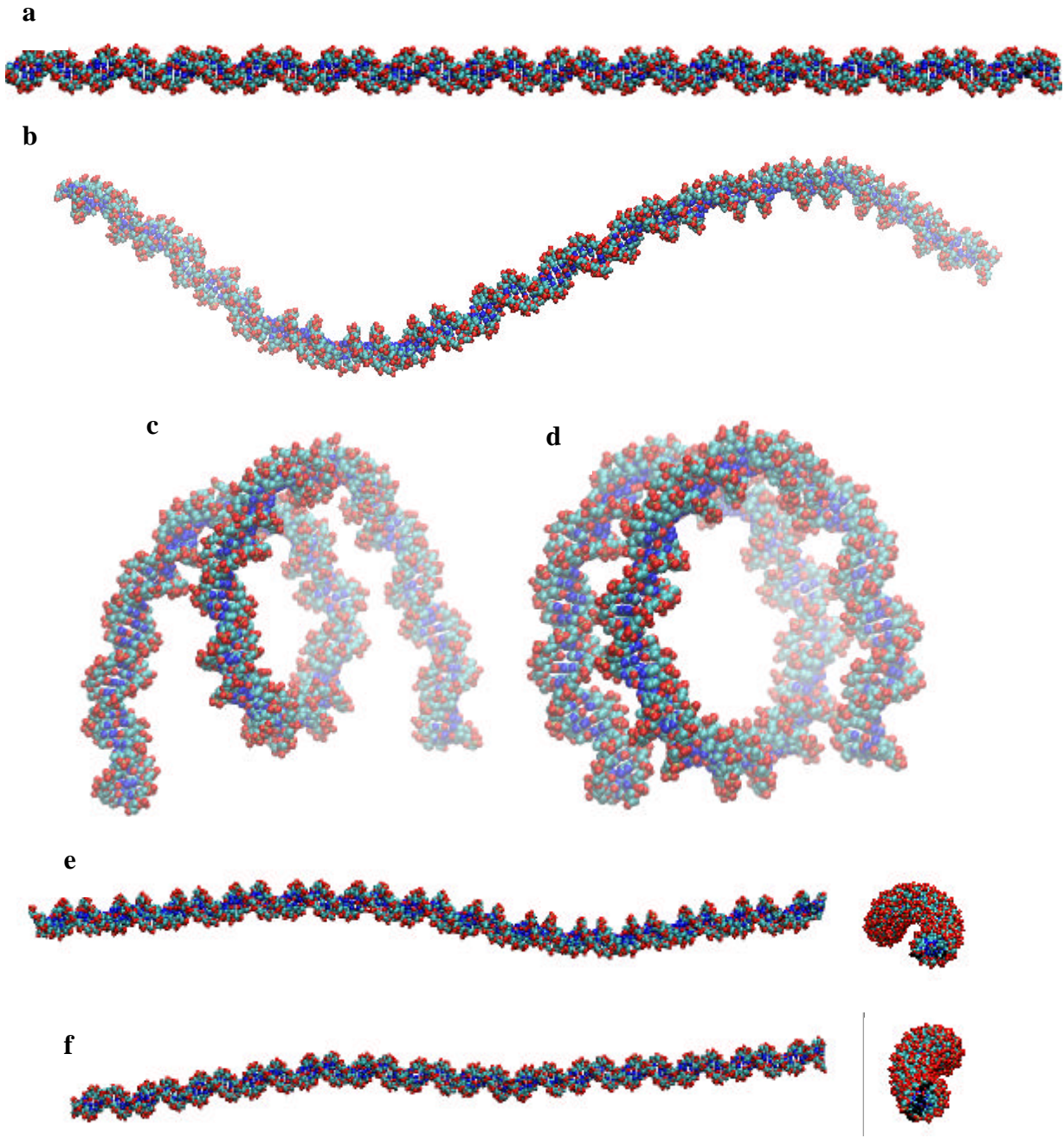

Figure 3. Reconstruction of DNA using selected Fourier Modes

**a)** – **d)** are models of *1kx3*. **e)** – **f)** are models of *dna*. **a)** Short wavelength modes [49-72] ~ 2.0-3.0bps. **b)** Intermediate and short wavelength modes [15-72] ~ 9.7-2.0bps. **c)** Modes [14-72] ~ 10.4-2.0bps. **d)** All modes [1-72] ~ 72.5-2.0bps. **e)** Intermediate wavelength modes [15-48] ~ 9.7-2.0bps, create a left-handed superhelix, far right image. **f)** Long wavelength modes [1-14] ~ 72.5-10.4bps create a right-handed superhelix, far right image. Depth queuing has been enabled in some images. Scales differ between models. All images created with VMD.